\documentclass{emulateapj}
\usepackage{epsfig}
\usepackage{amsmath}
\usepackage{amssymb}
\usepackage{natbib}
\usepackage{setspace}
\usepackage{graphicx}
\begin{document}
\title{The S$^4$G perspective on circumstellar dust extinction of AGB stars in M100}
%The S4G view of luminous red objects 
\author{Sharon E. Meidt\altaffilmark{1},
Eva Schinnerer\altaffilmark{1}, Juan-Carlos Mu\~noz-Mateos\altaffilmark{2}, Benne Holwerda\altaffilmark{3}, Luis C. Ho\altaffilmark{4}, Barry F. Madore\altaffilmark{4}, Johan H. Knapen\altaffilmark{5,6}, 
Albert Bosma\altaffilmark{7}, E. Athanassoula\altaffilmark{7}, Joannah L. Hinz\altaffilmark{8}, Kartik Sheth\altaffilmark{4,9,10}, Michael Regan\altaffilmark{11}, Armando Gil de Paz\altaffilmark{12}, Kar\'{i}n Men\'{e}ndez-Delmestre\altaffilmark{4}, Mark Seibert\altaffilmark{4}, Taehyun Kim\altaffilmark{2}, Trisha Mizusawa\altaffilmark{9,10}, Dimitri A. Gadotti\altaffilmark{13}, Eija Laurikainen\altaffilmark{14,15}, Heikki Salo\altaffilmark{14}, Jarkko Laine \altaffilmark{14,15}, S\'{e}bastien Comer\'{o}n\altaffilmark{16} }
\altaffiltext{1}{Max-Planck-Institut f\"ur Astronomie / K\"{o}nigstuhl 17 D-69117 Heidelberg, Germany}
\altaffiltext{2}{National Radio Astronomy Observatory}
\altaffiltext{3}{European Space Agency Research Fellow (ESTEC)}
\altaffiltext{4}{The Observatories of the Carnegie Institution for Science}
\altaffiltext{5}{Instituto de Astrofisica de Canarias, Spain}
\altaffiltext{6}{Departamento de Astrof\'\i sica, Universidad de La Laguna, Spain}
\altaffiltext{7}{Laboratoire d'Astrophysique de Marseille (LAM)}
\altaffiltext{8}{University of Arizona}
\altaffiltext{9}{Spitzer Science Center}
\altaffiltext{10}{California Institute of Technology}
\altaffiltext{11}{Space Telescope Science Institute}
\altaffiltext{12}{Universidad Complutense Madrid}
\altaffiltext{13}{European Southern Observatory}
\altaffiltext{14}{University of Oulu, Finland}
\altaffiltext{15}{Finnish Centre for Astronomy with ESO (FINCA), University of Turku}
\altaffiltext{16}{Korea Astronomy and Space Science Institute}
\date{\today}
%%%%%%%%%%%%%%%%%%%%%%%%%%%%%%\doublespacing
\begin{abstract}
We examine the effect of circumstellar dust extinction on the near-IR contribution of asymptotic giant branch (AGB) stars in intermediate-age clusters throughout the disk of M100.   
For our sample of 17 AGB-dominated clusters we extract optical-to-mid-IR SEDs and find that NIR brightness is coupled to the mid-IR dust emission in such a way that a significant reduction of AGB light, of up to 1 mag in K-band, follows from extinction by the dust shell formed during this stage.  
Since the dust optical depth varies with AGB chemistry (C-rich or O-rich), our results suggest that the contribution of AGB stars to the flux from their host clusters will be closely linked to the metallicity and the progenitor mass of the AGB star, to which dust chemistry and mass-loss rate are sensitive.
Our sample of clusters--each the analogue of a $\sim$1 Gyr old post-starburst galaxy--has implications within the context of mass and age estimation via SED modelling at high z: 
we find that the average $\sim$0.5 mag extinction estimated here may be sufficient to reduce the AGB contribution in (rest-frame) K-band from $\sim$70\%, as predicted in the latest generation of synthesis models, to $\sim$35\%.  Our technique for selecting AGB-dominated clusters in nearby galaxies promises to be effective for discriminating the uncertainties associated with AGB stars in intermediate-age populations that plague age and mass estimation in high-z galaxies. 
\end{abstract}
\section{Introduction}
\label{sec:intro}
Light emitted from stars in the thermally-pulsing asymptotic giant branch (AGB) phase of stellar evolution has implications for our view of the baryonic content and chemical state of galaxies, both locally and at high redshift, from UV-to-mid-IR observations.   
Recent advances in modeling and observation have enhanced this perspective, especially regarding the dust formation and mass loss that accompanies the late (thermal pulsing) stages of AGB evolution (e.g. \citealt{groenewegen}; \citealt{jackson}; \citealt{groenewegenDUST}; \citealt{boyer}).  These findings demonstrate the influential role AGBs play in the enrichment of the ISM with gas and dust (e.g. \citealt{srini}; \citealt{matsuura}; \citealt{sargent}) and place more accurate constraints on AGB lifetimes (i.e. \citealt{girardimarigo}). 

However, our view of stellar populations as a whole has only slowly incorporated this picture, and uncertainties in the ages and masses of intemediate-age systems remain. Because they are some of the brightest objects between ages 0.2-1 Gyr, a handful of $\mbox{AGB stars}$ can outshine all other stars at rest-frame wavelengths $\lambda\gtrsim1\mu m$ while contributing only minimally to the total mass.  Accurate AGB models are therefore a critical ingredient in stellar population synthesis (SPS) techniques (\citealt{ml1}; \citealt{bc03}; \citealt{maraston05}; Charlot \& Bruzual 2007; \citealt{gl}).  
But, even the latest generation of SPS models with an updated treatment of the AGB phase (e.g. Charlot \& Bruzual 2007 with \citealt{marigo}) rarely extend beyond $\mbox{3 $\mu m$}$, and few include the effect of extinction by the AGB star's circumstellar dust shell on the integrated properties of the stellar population (but see \citealt{mouhcine}).  
 
Here we examine whether recent evidence in favor of a reduced NIR AGB contribution \citep{kriek} arises from the extinction of AGBs by their dusty envelopes.  With a view extending from the optical to the mid-IR, we study a sample of AGB-dominated clusters in M100 selected for, and classified by, their dust emission.  
This avoids the challenges of optical or NIR detection and classification in the presence of the dust shell (e.g. \citealt{vanloon}) and supplies unique leverage on the role of dust to shift AGB light from the NIR to longer wavelengths.   
\section{The Data}
\label{sec:data}
We construct our sample of bright clusters in the disk of M100 (D=15.2 Mpc; \citealt{freedman}), chosen for its low inclination and the potential to probe a range of metallicities. Objects are selected based on their emission in IRAC 3.6 and $\mbox{4.5 $\mu m$}$ images processed by the S$^4$G team (\citealt{sheth}; see selection technique below).  In order to sample the SED from 0.2 to $\mbox{24 $\mu m$}$, we supplement these data with archival BVRI data from SINGS (\citealt{dale1}; \citealt{dale2}) and SINGS IRAC 5.8 and $\mbox{8 $\mu m$}$ and MIPS $\mbox{24 $\mu m$}$ images \citep{kennSings}.  
We also include \mbox{HAWK-I} JHK imaging provided by P. Grosbol (e.g. \citealt{grosbol}).  A map of continuum-subtracted H$\alpha$ from SINGS (\citealt{kennSings}; \citealt{calzetti}) provides cross-check against HII regions/young embedded clusters.   

Our cluster sample spans a large radial range in the disk, out to $R$$\sim$200'' (limited by the NIR field of view).   
Extrapolation of the measured gradient in the nebular emission-line chemical abundances \citep{moustakas} in M100 suggests that the ISM metallicity reaches 12+log(O/H)$\approx$8.1 by $R$=300''.  Assuming little chemical evolution over the last $\sim$1 Gyr, this suggests that the cluster sample spans 0.27$<$$Z/Z_{\odot}$$<$0.6.  In what follows, we refer to radial position and metallicity synonymously.   
\subsection{Selection of cluster candidates\label{sec:selection}}
The identification of red clusters at 3.6 and $\mbox{4.5 $\mu m$}$ is based on the Independent Component Analysis (ICA) of the images at these wavelengths (see \citealt{meidt11}).  ICA separates the light from the oldest stars, with colors $[3.6]-[4.5]$$<$0, from the emission contributed by contaminants, with colors 0$<$$[3.6]-[4.5]$$\lesssim$1.5.  Candidate bright red clusters, which fall into the second category, are distinguishable from the other main sources of contaminant emission at $\mbox{3.6 $\mu m$}$ given the very different relation to the non-stellar emission $F_{8,ns}$ at $\mbox{8 $\mu m$}$.     

\begin{figure}
\includegraphics[width=.85\linewidth]{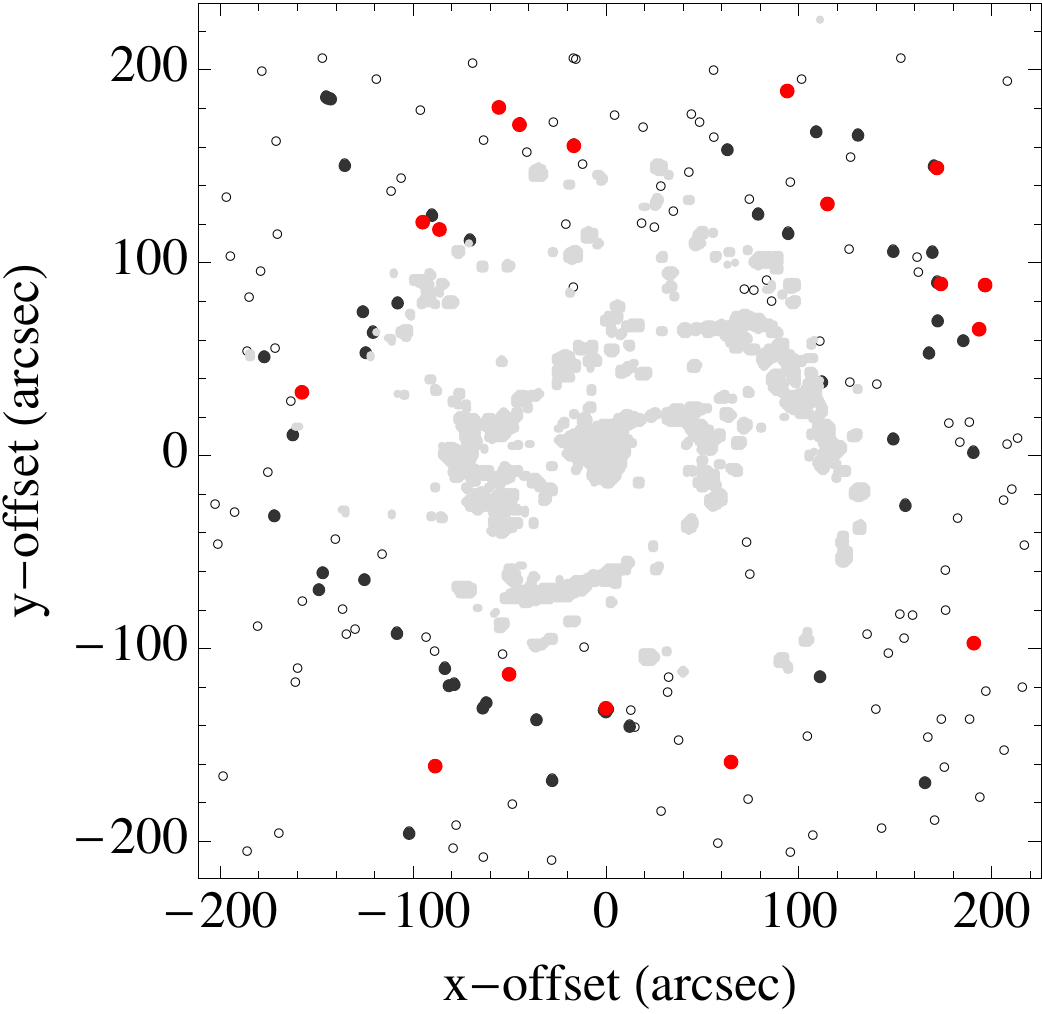}
\caption{Spatial map of candidate clusters (open symbols), significant detections at $\lambda\geq1.2 \mu m$ (black) and the final sample (red), together with the hot dust/PAH emission (gray), separated as described in the text.  Notice that PAH and hot dust emission more often appear in the high-density, actively star-forming regions.  \label{fig-map}}
\end{figure}

Using the selection criterion $F_{3.6}/F_{8,ns}$$>$0.3 to avoid emission from PAH and hot dust as advocated by \citet{meidt11}, we generate a map of 172 candidate clusters, shown in Figure \ref{fig-map}.  This map is used to identify the cluster emission in all wavebands, including at 3.6 and $\mbox{4.5 $\mu m$}$.  

\subsection{Multiband Photometry\label{sec:photometry}}
Photometric measurements for candidate clusters are extracted in circular apertures 6'' in radius, sized for compatibility with the 6'' PSF FWHM at $\mbox{24 $\mu m$}$, our lowest angular resolution image.  For this aperture size, sampling a region 444 pc in radius at the distance of this galaxy, we adopt corrections to the infinite aperture of 1.05, 1.07, 1.08, 1.09 and 1.67 at 3.6, 4.5, 5.8, 8 and $\mbox{24 $\mu m$}$, respectively, for point sources (SSC IRAC Handbook Handbook; Reach et al. 2005) and assume that these corrections are small or negligible in the optical images (i.e. \citealt{dale2}; \citealt{calzetti}).  We have confirmed that smaller apertures yield similar flux measurements for $\lambda$$<$$\mbox{24 $\mu m$}$.   

To isolate the cluster light from the underlying disk in each aperture we subtract local background values measured in annuli spanning between $r$=6'' and 12'' with 2-$\sigma_0$ clipping to avoid contamination by neighboring clusters.    
The variance in the local background value $\sigma_0$ defines the uncertainty in the photometric measurement $\sigma$ for each aperture, along with calibration and aperture correction uncertainties (added in quadrature).  Calibration uncertainties are $\sim$10\% in the NIR bands and below this level in the IRAC bands, at MIPS $\mbox{24 $\mu m$}$ and in the optical \citep{dale1}, while aperture corrections are typically 10\% uncertain in IRAC bands and $<$5\% uncertain at $\mbox{24 $\mu m$}$. 
\section{Photometric Properties}
\label{sec:props}
\subsection{Sample definition\label{sec:confirmation}}
Roughly half of the candidate clusters are detected in all bands longward of $\mbox{1 $\mu m$}$.   
Obscuration by dust presumably contributes to some non-detections in the NIR.  To ensure a realistic signal we adopt thresholding at 3$\sigma$.  Measurement uncertainties for the remaining 80 significant detections are well below 20\% (nearer 10\% on average) and are omitted for the final sample in all upcoming plots for clarity.  

Candidate clusters have some of the reddest colors in the shortest IRAC wavebands (0$<$$\mbox{[3.6]-[4.5]}$$<$1; \citealt{meidt11}) consistent with the colors of $\mbox{AGB stars}$.  Still, possible contaminants include background galaxies, quasars, and younger dust enshrouded clusters.   Galaxy number counts over the range in $\mbox{3.6 $\mu m$}$ magnitudes in our sample are high (e.g. \citealt{fazio}, \citealt{sanders}), but only a subset have $\mbox{[3.6]-[4.5] colors}$ that can be confused with emission from dusty AGBs, namely quasars and the nearest and brightest background galaxies, for which the observed mid-IR colors are closest to rest-frame colors (e.g. \citealt{fazio}).  In the latter case, 14 are removed with a minimum size criterion (set by the $\mbox{3.6 $\mu m$}$ FWHM), and 12 others avoided with limits to the mid-IR colors: star-forming galaxies will have [3.6]-[8]$>$3.35 (tracing hot dust and PAH emission) and [8]-[24]$\gtrsim$2.4 (see \citealt{boyer}).

Contamination from background galaxies is otherwise naturally limited by the optically thick disk of M100 \citep{holwerda}.  As a secondary safeguard we use the few known background galaxies detected by \citet{holwerda} between 70'' and 140'' in archival HST/WFPC2 imaging to construct a composite SED against which all candidates are compared.  These galaxies are on average one magnitude dimmer at $\mbox{3.6 $\mu m$}$ than clusters in our final sample and have B-K$\gtrsim$3, K-[8]$\gtrsim$4 and [3.6]-[8.0]$\gtrsim$3.  
Note that sources falling nearest the disk edge where the dust column density is lower show less strongly the suppression at optical wavelengths characteristic of foreground disk extinction.  We therefore expect the above mid-IR color criteria to be a stronger limit for these.  

\begin{figure*}[rt]
\begin{tabular}{cc}
\includegraphics[width=.45\linewidth]{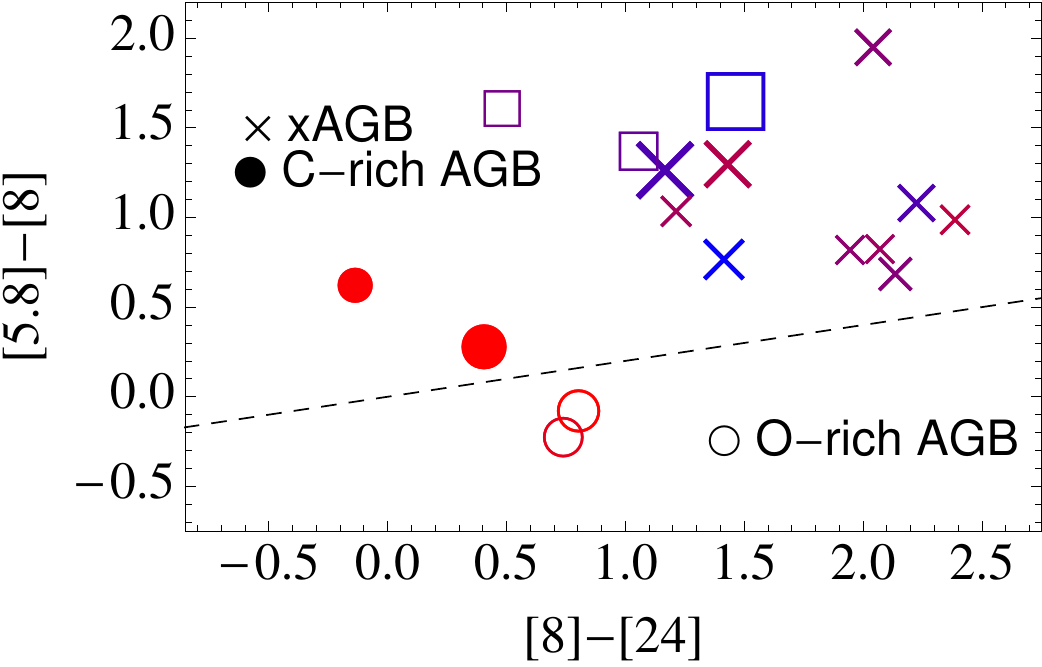}&\includegraphics[width=.45\linewidth]{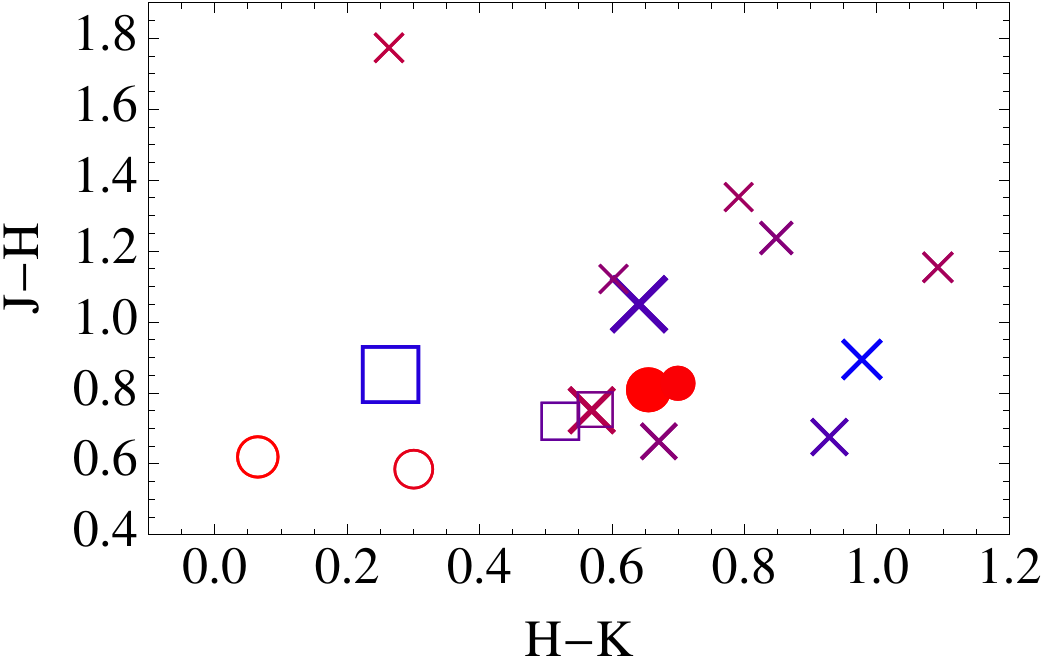}\\
\includegraphics[width=.425\linewidth]{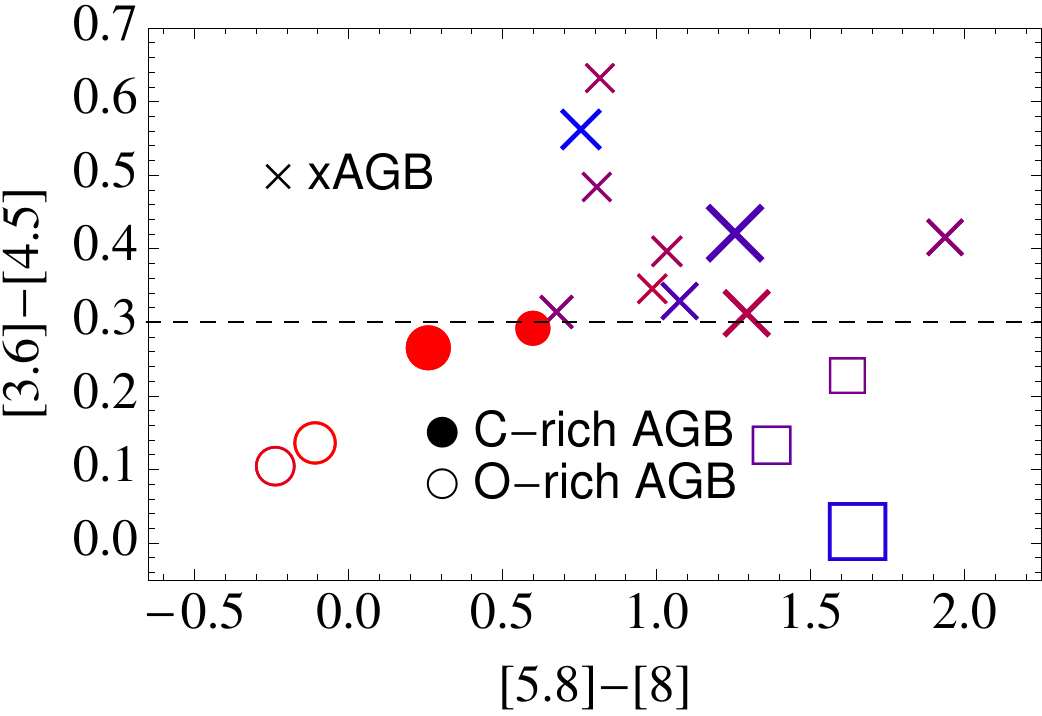}&\includegraphics[width=.45\linewidth]{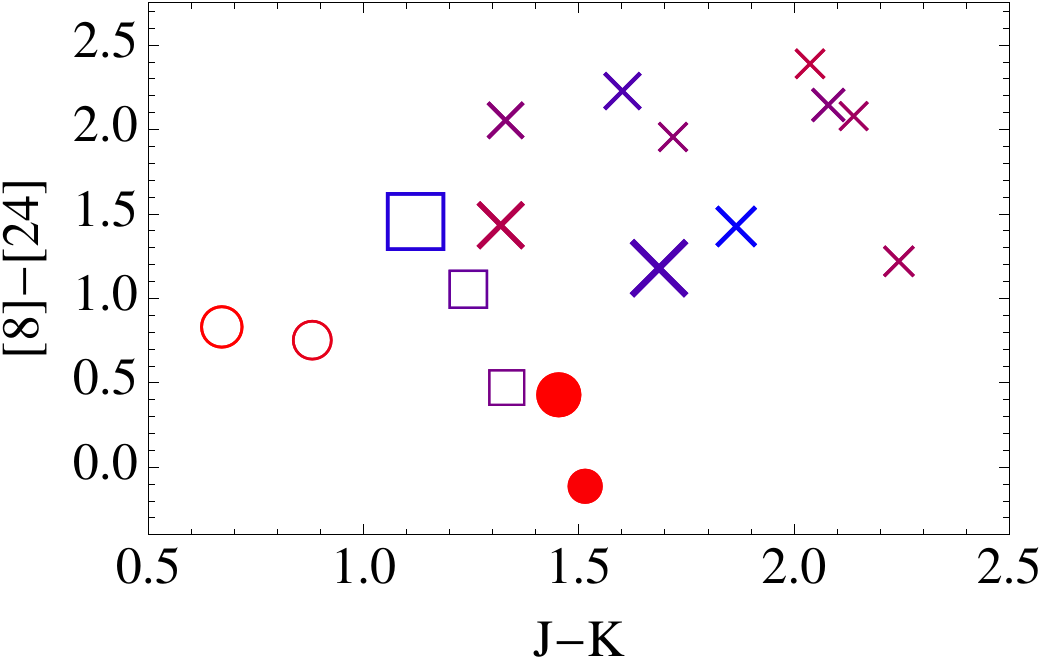}\\
\includegraphics[width=.425\linewidth]{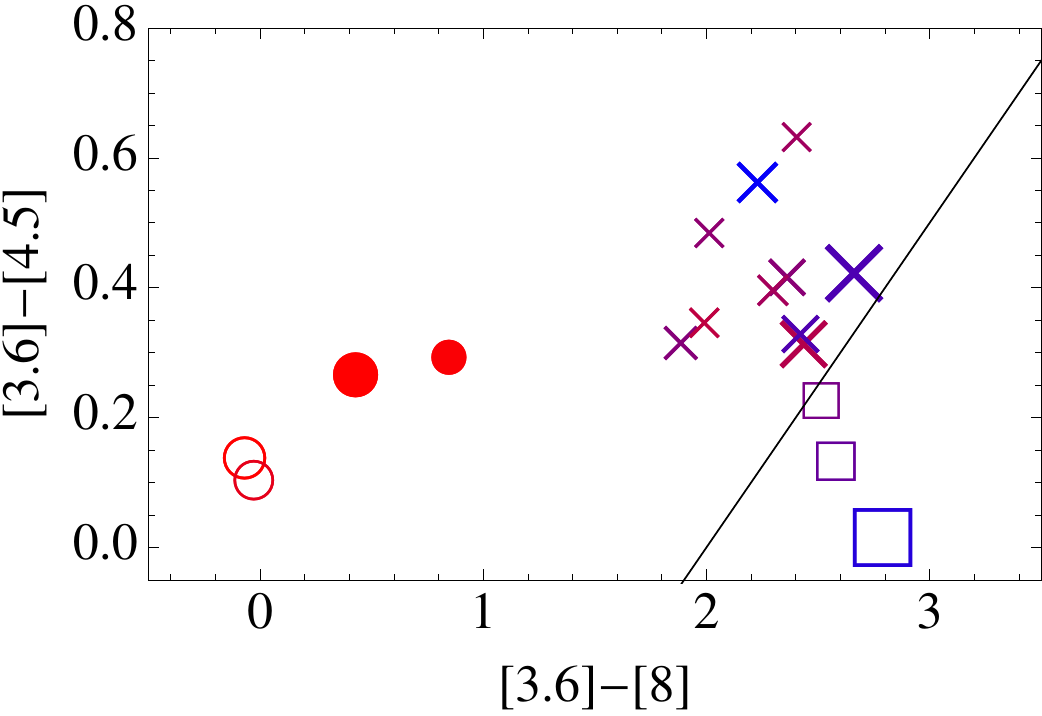}&\includegraphics[width=.425\linewidth]{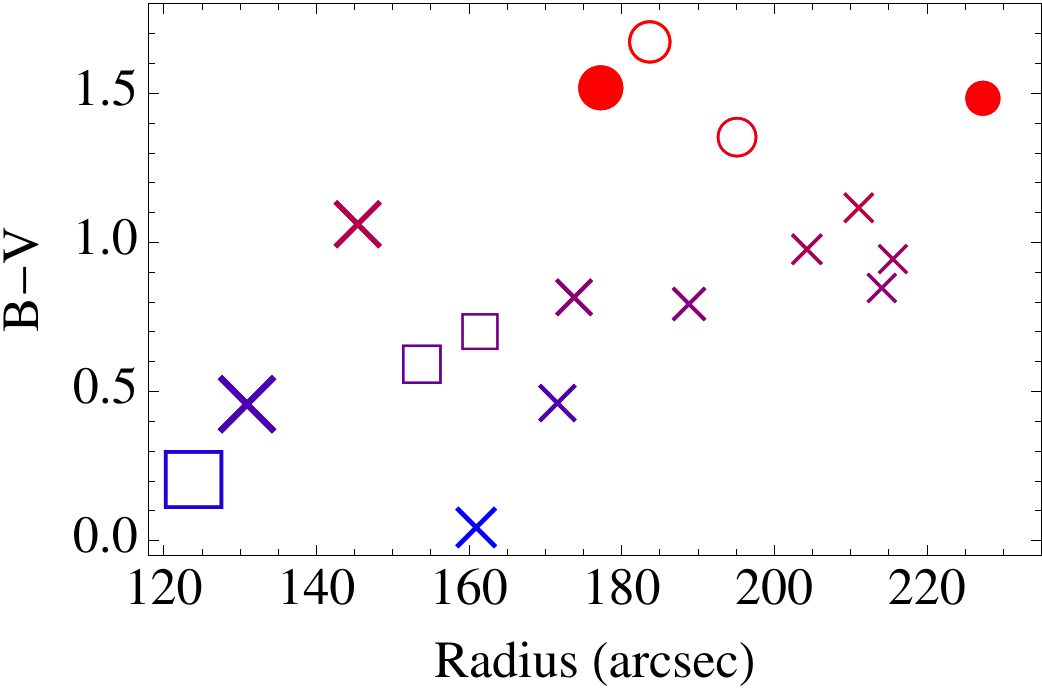}\\
\end{tabular}
\caption{{\it (Left Column)} Separation between clusters hosting C-rich (closed), O-rich (open) or extreme (crosses) $\mbox{AGB stars}$.  Dashed gray lines  show our criteria for distinguishing between the three main types, as motivated by \citet{boyer}.  Square symbols mark a fourth subset of `mid-IR bright' objects to the left of the solid black line in the bottom panel, which could potentially be dusty young clusters (e.g. \citealt{corbelli}), later classified in this work as anomalously dusty O-rich stars (see Figure \ref{fig-cosep}).  The color scaling represents variation in B-V~color, from B-V=0 (blue) to B-V=1.5 (red), used as a proxy for cluster age, 
while the symbol size varies according to metallicity, from low (small) to high (large).  {\it (Right Column)}({\it Top}) J-H vs. $\mbox{H-K color}$ for all objects in the sample.  ({\it Middle}) [8]-[24] vs. J-K diagram as used by \citet{boyer} to distinguish between dust chemistries.  The aO-rich AGBs with silicate dust follow a nearly vertical sequence near J-K$\sim$1.3, while extreme C-rich AGBs are distributed more horizontally.  ({\it Bottom}) Positions and $\mbox{B-V colors}$ of our clusters tracking metallicity and cluster age, respectively.  
\label{fig-cosep}}
\end{figure*}

According to the detection rate of AGN behind the LMC and SMC we expect quasar contamination to be minimal (less than 2 in the field studied here; \citealt{kkquasars}), and few candidates approach the limit R-K$\sim$4 used to detect obscured quasars with similar mid-IR colors (e.g. \citealt{lacy}; \citealt{gilkman}).  
Embedded star formation in young massive clusters and HII regions illuminating the surrounding dust are perhaps stronger potential contaminants.  
Our imposed criterion [3.6]-[8]$<$3.35 already avoids many such objects, 4 of which also have detectable H$\alpha$ emission.  To select against any more deeply embedded mid-IR bright clusters we impose a 3$\sigma$ threshold in the optical BVRI bands.   With 27 such sources removed, the remaining 17 visually confirmed clusters exhibit 
 $\mbox{[3.6]-[8] colors}$ consistent with emission from circumstellar material (see Figure \ref{fig-cosep}), as expected around $\mbox{AGB stars}$ undergoing high mass loss (e.g. \citealt{groenewegenDUST}; \citealt{sargent1}; \citealt{srini11}).  

\subsection{Sample description}
By construction, selected clusters originate in zones with little to no emission from PAH and hot dust at 3.6 and $\mbox{4.5 $\mu m$}$ and so they should also be largely removed of locations in the disk susceptible to the effects of reddening and extinction by diffuse dust (see Figure \ref{fig-map}).  Extra-cluster extinction is therefore expected to be minimal, especially since the final 17 clusters fall in the outer disk where the dust column density is low.  As revealed in $\S$ \ref{sec:co}, our mid-IR measurements are consistent with emission from the circumstellar dust shell produced, and illuminated, by the $\mbox{AGB star}$, itself, rather than from a diffuse dust component emitting re-processed radiation from cluster stars.   We 
take the measured red $\mbox{B-V colors}$ (Figure \ref{fig-cosep}) as characteristic of intrinsically red objects with a spread in ages between 0.2-1 Gyr (\citealt{b07}; \citealt{ml1}).  

Our sample is evidently composed of coeval massive cluster complexes with M$>$10$^6$ M$_\odot$ (e.g. \citealt{trancho}; referred to simply as clusters in what follows), given the unavoidable bias towards the brightest sources at this distance and our requirement for optical detections.  The depth of the potential of the cluster complex relative to the shallow outer disk provides a natural way to avoid dispersal at these late ages (i.e. \citealt{fellhauer}), while interaction with the companion galaxy suggests a plausible seed for the formation of such massive complexes (e.g. \citealt{larsen}).  
\section{Variation in the AGB Contribution}
\label{sec:env}
\subsection{Distinction between O-rich, C-rich and extreme AGB stars\label{sec:co}}
The varying degrees of circumstellar dust extinction expected around AGBs are potentially closely related to the chemistry of the star and the mass of its progenitor.  
C-rich stars will be more highly extincted than O-rich stars for the same mass-loss rate, given the higher opacities of the carbonaceous grains produced in their atmospheres (e.g. \citealt{suh}; \citealt{pegorie}; \citealt{mouhcine}).   The highest levels of extinction occur in the extreme AGB phase, during which the star undergoes its highest mass-loss rates.  
Following \citet{boyer}, we use mid-IR colors to differentiate between clusters dominated by C-rich, O-rich or extreme $\mbox{AGB stars}$ (Figure \ref{fig-cosep}).  

Although age and metallicity variations between clusters may introduce scatter relative to the separation in, e.g., the LMC \citep{boyer}, 
the different $\mbox{AGB star}$ types cleanly separate by their mid-IR colors.    Note that, without spectral confirmation, these particular classifications are subject to interpretation.  Following \citet{kastner} and \citet{groenewegenDUST}, for example, the xAGBs here would be classified as C-rich (and see \citealt{ferraro}).  

Clusters also show good correspondence with the $\mbox{NIR colors}$ of isolated AGBs (top right panels of Figure \ref{fig-cosep}), despite the  higher contribution from the rest of the cluster members than in the mid-IR.  These very red NIR colors supply additional confirmation that our clusters are dominated by AGBs undergoing high mass-loss rates (e.g. \citealt{gl}); even with the implied color excess E(B-V) for ages between 1-10 Myr, we find that the NIR colors remain consistent with 0.2-1.2 Gyr old clusters.   

Our classifications are confirmed in the [8]-[24] vs. J-K diagram (middle; cf. \citealt{boyer}).   Based on this diagram our `mid-IR bright' objects can be associated with the anomalously dusty O-rich (aO-rich) AGBs identified by \cite{boyer}.   This is also favored by our combined optical and mid-IR view: where the red $\mbox{[3.6]-[8] colors}$ of dust-reprocessed emission from young embedded clusters would coincide with dust-reddened $\mbox{B-V colors}$, these objects have some of the bluest optical colors in the sample.  These objects may therefore be the O-rich counterpart to the extreme C-rich AGBs in our sample.  

The bottom right panel of Figure \ref{fig-cosep} depicts the dependence of AGB type on mass and metallicity, in line with the mechanisms that influence the C/O ratio (see, e.g., \citealt{vanloon}).
The dustiest AGBs notably sit in younger clusters, where AGB progenitors are of higher mass than at older ages.  
This accounts for the enhanced dust production in these stars, if the implied longer pulsation period for these high masses leads to higher mass-loss rates.  In contrast, the less dusty C-rich and O-rich appear less often here and only at the outer radii as a result of our mid-IR and optical detection thresholds and large aperture.  \subsection{Extinction in $\mbox{K band}$\label{sec:reduction}}
\begin{figure}[t]
\includegraphics[width=.85\linewidth]{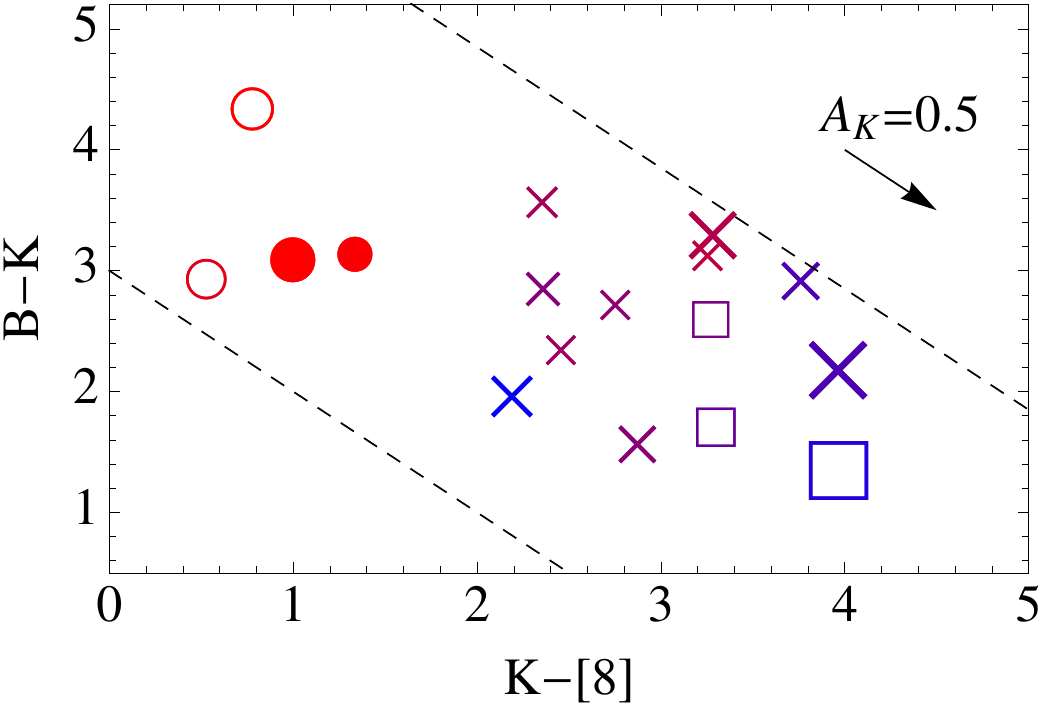}
\caption{B-K vs. $\mbox{K-[8] color}$ for the final cluster sample.  Dashed lines illustrate decreasing K-band brightness at fixed B and $\mbox{8 $\mu m$}$ magnitudes.  Symbol types, sizes  and colors are as in Figure \ref{fig-cosep}.  \\}
\label{fig-8k}
\end{figure}

The impact of the different $\mbox{AGB star}$ dust chemistries and mass-loss rates is clear in Figure \ref{fig-8k} where clusters form a sequence that becomes progressively redder in $\mbox{K-[8] color}$ while becoming bluer in $\mbox{B-K color}$.  
  The optical-NIR color traces the fractional flux contribution of $\mbox{AGB stars}$ relative to the other (less evolved) cluster members, which dominate in the optical.  The K-[8]~color meanwhile serves to measure the amount of dust produced around the $\mbox{AGB stars}$, in the sense that enhanced $\mbox{8 $\mu m$}$ dust emission and reduced K-band brightness due to extinction both drive reddening in K-[8].

To reproduce the downward trend in this color-color diagram that separates the C-rich from the O-rich AGBs and the dustiest AGBs from those that are less dusty requires a decrease in the K-band cluster light from the top left to the bottom right (depicted by the overlaid dashed lines of K-band dimming at arbitrary fixed B and [8] brightness).  This can be most easily explained by an enhanced obscuration of the $\mbox{AGB stars}$ as the optical depth of their dust shells increase.  

The stellar photosphere or spectral type of the $\mbox{AGB star}$ also plays a role in determining the location within this diagram (e.g, \citealt{sargent1}).   This is contained in our distinction between C-rich and O-rich chemistries and so, in this sense, it is largely indistinguishable from the implied dust extinction.  As a result of the different properties of the dust grains produced in the atmospheres of C-rich and O-rich stars, the former will be relatively more obscured for the same mass-loss rate.  

Other interpretations for the downward trend in Figure \ref{fig-8k} are either less viable or precluded by construction. By avoiding sources with mid-IR emission from dust heated by young stars we have removed the potential that the bluer B-K colors arise at an age before the onset of AGBs, as well as the possibility that reddest K-[8] colors trace extracluster dust heated to higher temperatures.  

In addition, the decrease in B-K is inconsistent with fluctuations to lower $\mbox{AGB star}$ number densities (resulting in lower NIR fluxes relative to the optical), as the corners of the diagram would also be filled, since B-K would be independent of K-8 in this case.  

This trend is also not driven by intra-cluster dust from past mass loss by earlier AGBs at younger cluster ages, which would have reversed the observed red-to-blue progression in B-V colors along the downward sequence of increasing extinction.  Instead, clusters that are blue in B-K are also blue in B-V.  As noted in $\S$ \ref{sec:co}, this reflects a genuine dependence on cluster age through the mass of the progenitor; at young cluster ages AGBs will evolve from higher mass stars resulting in AGBs with higher mass-loss rates and hence heavier extinction.

\subsection{Implications for age and mass estimation\label{sec:implications}}
The trend in Figure \ref{fig-8k} implies that the contribution of dusty AGBs in K-band is much lower than expected from the B-band cluster emission due to circumstellar extinction.  
Since optical-NIR colors like B-K are commonly used as age indicators for intermediate-age populations (\citealt{b07}; \citealt{mouhcine}), this suggests that intrinsic extinction can significantly affect the predicted ages of AGB-dominated populations, as explored by \citet{mouhcine}.  Age estimation and classification from NIR colors should be likewise affected (e.g. \citealt{nowotny}). 
\begin{figure}[t]
\includegraphics[width=.9\linewidth]{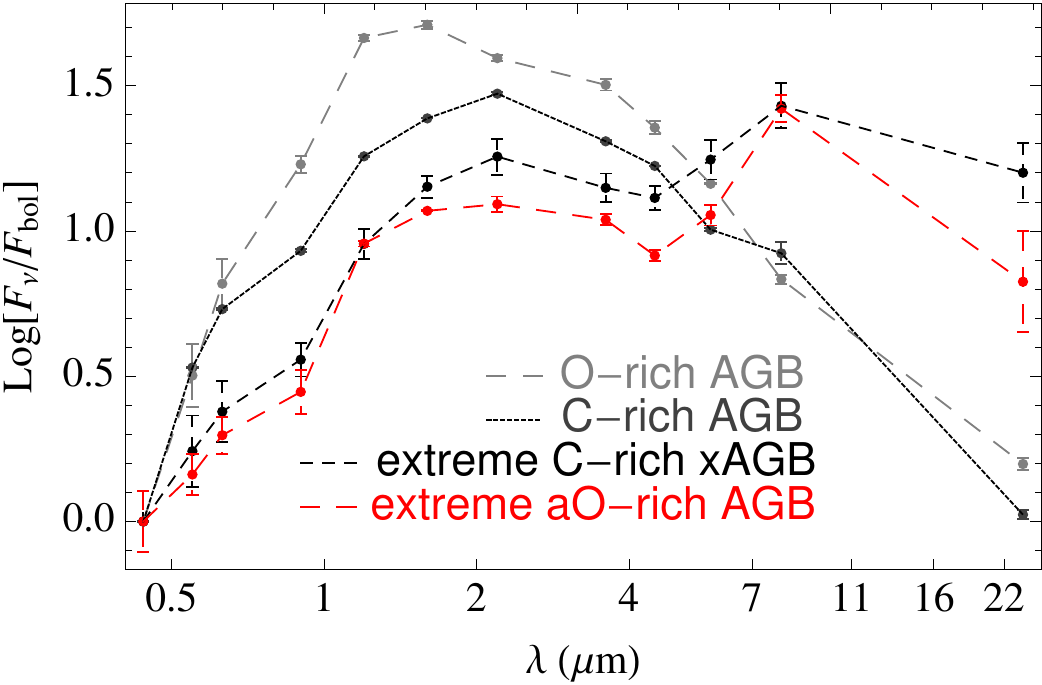}
\caption{Composite SEDs normalized at $\mbox{0.4 $\mu m$}$.  For a given type, we take the average of SEDs each normalized to its bolometric flux $F_{bol}$.  Error bars represent the dispersion in the measurements for each type. \label{fig-compseds}}
\end{figure}
More consequentially, models with relatively `bare' AGBs, less extincted by circumstellar dust, will underpredict the stellar M/L for intermediate-age populations.  This is demonstrated in Figure \ref{fig-compseds}, showing composite SEDs for the four sets of clusters studied here, those that host O-rich and C-rich AGBs and those with relatively more obscured extreme C-rich and aO-rich AGBs.  In the latter two cases, circumstellar dust clearly shifts the cluster emission from short to long wavelengths so that the NIR contribution of AGBs appears significantly lower than in the case of relatively `bare' O-rich $\mbox{AGBs}$.  The C-rich $\mbox{AGBs}$ also shows a depression in the NIR relative to the O-rich case.   This amounts to 0.12 dex $\approx$ 0.3 mag extinction in K-band for C-rich stars compared to O-rich stars and on average $\sim$0.5 dex extinction for the dustiest AGBs (or as much as 1 mag, also judging by the offset in B-K color in Figure \ref{fig-8k}).   

The difference in SED shapes with and without strong $\mbox{AGB star}$ extinction can explain the apparent compatibility found by \citet{kriek} between the SEDs of post-starburst galaxies and the older generation of \citet{bc03} models, where AGBs contribute $\sim$30-40\% in K at 1 Gyr rather than $\sim$70\% in the recent, updated Bruzual \& Charlot (2007) models.  
But the agreement found by \citet{kriek} would seem to be largely coincidental: taking the extinction measured from difference in K-band light in the undimmed and dimmed cases, then a 70\% contribution from $\mbox{AGB stars}$ in K-band (30\% from the cluster) will be reduced to $\sim$ 35\% as the result of circumstellar dust extinction.  

\section{Concluding Remarks}
We have presented evidence that the fractional contribution of AGBs to intermediate-age populations, a contentious prediction of stellar population synthesis models, is significantly affected by the presence of the dusty envelope produced during this stage.  The extinction generated by dust shells around AGBs in $\sim$1 Gyr old clusters (on average 0.5 mag) is sensitive to the type of $\mbox{AGB star}$: C-rich AGBs produce shells with higher optical depths than their O-rich counterparts and the high mass-loss rates of xAGBs lead to even higher obscuration. Consequently, population ages estimated from optical-NIR colors may be underestimated (as predicted by \citealt{mouhcine}) leading also to under-prediction of the stellar M/L.  The range in circumstellar extinction revealed here may account for the fact that SPS models with very different $\mbox{AGB star}$ contributions each seem to fit a diversity of observations well (cf. \citealt{maraston06}, \citealt{vdwel06} and \citealt{kriek}).\\

S.E.M. thanks Arjen van der Wel for discussion and Angela Adamo and Anibal Garcia for their helpful comments on the manuscript.  
E.A. and A.B. thank the Centre National d'Etudes Spatiales for financial support.
\newpage
%\begin{landscape}
\begin{table}\footnotesize
\begin{center}
\caption{Catalog of AGB-dominated clusters\protect\footnote{All photometric measurements are presented relative to Vega.}}

\begin{tabular}{l l l l l l l l l l l l l ll}
\tableline\tableline
ID&RA&DEC&$m_B$&$dm_B$&$m_V$&$dm_V$&$m_R$&$dm_R$&$m_I$&$dm_I$&$m_J$&$dm_J$&$m_H$&$dm_H$\\
&(deg)&(deg)&(mag)&(mag)&(mag)&(mag)&(mag)&(mag)&(mag)&(mag)&(mag)&(mag)&(mag)&(mag)\\

1& 185.7518 & 
  15.7765 & 21.158 & 0.103 & 19.503 & 0.101 & 17.987 & 
    0.100 & 16.842 & 0.100 & 14.918 & 0.050 & 14.312 & 
    0.050 \\

2&185.709 & 
  15.7771 & 22.275 & 0.250 & 21.813 & 0.278 & 20.462 & 
    0.178 & 19.914 & 0.162 & 18.350 & 0.080 & 17.677 & 
    0.084 \\

3& 185.727 & 
  15.7849 & 21.667 & 0.135 & 21.215 & 0.153 & 20.770 & 
    0.176 & 20.241 & 0.145 & 18.573 & 0.072 & 17.526 & 
    0.073 \\

4& 185.741 & 
  15.7898 & 20.616 & 0.156 & 20.405 & 0.182 & 20.211 & 
    0.266 & 20.543 & 0.313 & 17.764 & 0.098 & 16.910 & 
    0.095 \\

5& 185.674 & 
  15.7942 & 21.308 & 0.106 & 20.467 & 0.103 & 19.414 & 
    0.102 & 18.893 & 0.104 & 17.708 & 0.052 & 16.591 & 
    0.054 \\

6&185.771 & 
  15.8304 & 21.060 & 0.140 & 21.022 & 0.183 & 20.639 & 
    0.226 & 20.222 & 0.178 & 18.366 & 0.112 & 17.478 & 
    0.097 \\

7&185.674 & 
  15.8395 & 22.149 & 0.120 & 21.175 & 0.110 & 19.951 & 
    0.108 & 20.044 & 0.119 & 18.217 & 0.054 & 17.065 & 
    0.064 \\

8&185.679 & 
  15.846 & 19.739 & 0.100 & 18.393 & 0.100 & 17.151 & 
    0.100 & 16.603 & 0.100 & 15.093 & 0.050 & 14.512 & 
    0.050 \\

9&185.673 & 
  15.8458 & 20.974 & 0.103 & 20.036 & 0.102 & 19.213 & 
    0.102 & 19.477 & 0.112 & 18.165 & 0.055 & 16.818 & 
    0.056 \\

10&185.751 & 
  15.8538 & 21.258 & 0.270 & 20.197 & 0.167 & 19.199 & 
    0.140 & 18.725 & 0.115 & 16.672 & 0.060 & 15.922 & 
    0.060 \\

11&185.754 & 
  15.8549 & 20.137 & 0.190 & 19.544 & 0.174 & 18.787 & 
    0.158 & 18.884 & 0.158 & 17.061 & 0.074 & 16.339 & 
    0.074 \\

12&185.695 & 
  15.8575 & 20.330 & 0.106 & 19.520 & 0.106 & 18.797 & 
    0.106 & 18.923 & 0.108 & 17.497 & 0.061 & 16.837 & 
    0.062 \\

13&185.68 & 
  15.8627 & 20.242 & 0.101 & 18.773 & 0.100 & 17.740 & 
    0.100 & 17.500 & 0.101 & 16.041 & 0.050 & 15.226 & 
    0.050 \\

14&185.732 & 
  15.8659 & 19.969 & 0.158 & 19.262 & 0.138 & 18.163 & 
    0.117 & 17.887 & 0.109 & 16.074 & 0.053 & 15.316 & 
    0.053 \\

15&185.74 & 
  15.8689 & 20.281 & 0.107 & 18.773 & 0.101& 17.738 & 
    0.101 & 17.481 & 0.100 & 16.060 & 0.050 & 15.262 & 
    0.050 \\

16&185.743 & 
  15.8714 & 21.781 & 0.111 & 20.994 & 0.111 & 19.964 & 
    0.108 & 20.352 & 0.126 & 18.411 & 0.057 & 17.180 & 
    0.063 \\

17&185.701 & 
  15.8738 & 22.956 & 0.151 & 21.840 & 0.126 & 21.192 & 
    0.138 & 20.157 & 0.105 & 19.253 & 0.069 & 17.479 & 
    0.058 \\
\end{tabular}\\
\begin{tabular}{l l l l l l l l l l l l l l l}
\tableline\tableline
ID&\hspace*{.82in}&\hspace{1cm}&$m_K$&$dm_K$&$m_{3.6}$&$dm_{3.6}$&$m_{4.5}$&$dm_{4.5}$&$m_{5.8}$&$dm_{5.8}$&$m_{8}$&$dm_{8}$&$m_{24}$&$dm_{24}$\\
&\hspace*{.82in}&&(mag)&(mag)&(mag)&(mag)&(mag)&(mag)&(mag)&(mag)&(mag)&(mag)&(mag)&(mag)\\
1&\hspace*{.82in}&&14.247 & 0.050 & 13.400 & 0.112 & 13.270 & 
    0.112 & 13.360 & 0.113 & 13.468 & 0.116 & 12.664 & 
    0.079\\
2&&&16.748 & 0.068 & 15.408 & 0.116 & 15.080 & 
    0.115 & 14.063 & 0.122 & 12.987 & 0.125 & 10.762 & 
    0.072\\
3&&&16.885 & 0.065 & 15.583 & 0.114 & 15.164 & 
    0.113 & 14.175 & 0.123 & 12.918 & 0.125 & 11.747 & 
    0.076\\
4&&&16.643 & 0.100 & 15.465 & 0.130 & 15.448 & 
    0.132 & 14.332 & 0.160 & 12.676 & 0.157 & 11.212 & 
    0.081\\
5&&&15.988 & 0.052 & 15.250 & 0.112 & 14.768 & 
    0.112 & 14.043 & 0.115 & 13.236 & 0.114 & 11.290 & 
    0.071\\
6&&&16.500 & 0.071 & 16.538 & 0.149 & 15.979 & 
    0.126 & 15.064 & 0.161 & 14.310 & 0.257 & 12.894 & 
    0.109\\
7&&&15.972 & 0.061 & 15.917 & 0.114 & 15.521 & 
    0.115 & 14.653 & 0.124 & 13.619 & 0.118 & 12.404 & 
    0.080\\
8&&&14.212 & 0.050 & 13.656 & 0.112 & 13.555 & 
    0.112 & 13.447 & 0.113 & 13.684 & 0.119 & 12.945 & 
    0.087\\
9&&&16.026 & 0.054 & 15.974 & 0.114 & 15.344 & 
    0.113 & 14.386 & 0.120 & 13.570 & 0.117 & 11.498 & 
    0.072\\
10&&&15.352 & 0.057 & 14.503 & 0.114 & 14.190 & 
    0.113 & 13.361 & 0.121 & 12.066 & 0.131 & 10.632 & 
    0.073\\
11&&&15.817 & 0.067 & 15.098 & 0.121 & 14.964 & 
    0.120 & 13.892 & 0.123 & 12.518 & 0.123 & 11.462 & 
    0.078\\
12&&&16.166 & 0.057 & 15.655 & 0.118 & 15.242 & 
    0.115 & 15.231 & 0.139 & 13.294 & 0.116 & 11.251 & 
    0.072\\
13&&&14.525 & 0.050 & 14.036 & 0.112 & 13.750 & 
    0.112 & 13.788 & 0.114 & 13.189 & 0.114 & 13.324 & 
    0.094\\
14&&&14.741 & 0.052 & 13.985 & 0.113 & 13.755 & 
    0.113 & 13.091 & 0.119 & 11.470 & 0.121 & 10.988 & 
    0.078\\
15&&&14.606 & 0.050 & 14.037 & 0.112 & 13.776 & 
    0.112 & 13.868 & 0.116 & 13.608 & 0.131 & 13.202 & 
    0.103\\
16&&&16.331 & 0.068 & 15.858 & 0.114 & 15.546 & 
    0.115 & 14.649 & 0.122 & 13.973 & 0.135 & 11.836 & 
    0.075\\
17&&&17.216 & 0.056 & 15.953 & 0.114 & 15.608 & 
    0.115 & 14.949 & 0.132 & 13.962 & 0.122 & 11.574 &
    0.072\\
\end{tabular}
\end{center}
\end{table}
%\end{landscape}

\end{document}